\documentclass[aps, prd, amsmath, floats, floatfix, twocolumn, nofootinbib, superscriptaddress]{revtex4-1}
\usepackage[pdftex]{graphicx}
\usepackage{color}
\usepackage{latexsym}
\usepackage[colorlinks]{hyperref}
\usepackage{amsmath,physics}

\newcommand{\be}{\begin{eqnarray}}
\newcommand{\ee}{\end{eqnarray}}

\newcommand{\bwide}{\begin{widetext}}
\newcommand{\ewide}{\end{widetext}}

\begin{document}

\title{Constraining the Konoplya-Rezzolla-Zhidenko deformation parameters II:\\limits from stellar-mass black hole X-ray data}

\author{Zhibo~Yu}
\affiliation{Center for Field Theory and Particle Physics and Department of Physics, Fudan University, 200438 Shanghai, China}

\author{Qunfeng~Jiang}
\affiliation{Center for Field Theory and Particle Physics and Department of Physics, Fudan University, 200438 Shanghai, China}

\author{Askar~B.~Abdikamalov}
\affiliation{Center for Field Theory and Particle Physics and Department of Physics, Fudan University, 200438 Shanghai, China}
\affiliation{Ulugh Beg Astronomical Institute, Tashkent 100052, Uzbekistan}

\author{Dimitry~Ayzenberg}
\affiliation{Theoretical Astrophysics, Eberhard-Karls Universit\"at T\"ubingen, D-72076 T\"ubingen, Germany}

\author{Cosimo~Bambi}
\email[Corresponding author: ]{bambi@fudan.edu.cn}
\affiliation{Center for Field Theory and Particle Physics and Department of Physics, Fudan University, 200438 Shanghai, China}

\author{Honghui~Liu}
\affiliation{Center for Field Theory and Particle Physics and Department of Physics, Fudan University, 200438 Shanghai, China}

\author{Sourabh~Nampalliwar}
\affiliation{Theoretical Astrophysics, Eberhard-Karls Universit\"at T\"ubingen, D-72076 T\"ubingen, Germany}

\author{Ashutosh~Tripathi}
\affiliation{Center for Field Theory and Particle Physics and Department of Physics, Fudan University, 200438 Shanghai, China}

\begin{abstract}
Astrophysical black holes are thought to be the Kerr black holes predicted by general relativity, but macroscopic deviations from the Kerr solution can be expected from a number of scenarios involving new physics. In Paper~I, we studied the reflection features in \textsl{NuSTAR} and \textsl{XMM-Newton} spectra of the supermassive black hole at the center of the galaxy MCG--06--30--15 and we constrained a set of deformation parameters proposed by Konoplya, Rezzolla \& Zhidenko (Phys. Rev. D93, 064015, 2016). In the present work, we analyze the X-ray data of a stellar-mass black hole within the same theoretical framework in order to probe a different curvature regime. We consider a \textsl{NuSTAR} observation of the X-ray binary EXO~1846--031 during its outburst in 2019. As in the case of Paper~I, all our fits are consistent with the Kerr black hole hypothesis, but some deformation parameters cannot be constrained well. 
\end{abstract}

\maketitle


\section{Introduction}

X-Ray reflection spectroscopy (XRS) is a powerful tool for studying the properties of black holes (BHs) and the accretion disks that surround them~\cite{Reynolds:2013qqa,Bambi:2017iyh,Bambi:2020jpe}. The reflection spectrum of a BH accretion disk is generated through a multi-step process. The accretion disk emits thermal radiation in the form of a multi-temperature blackbody spectrum, some of which impinges on a hotter corona somewhere in the vicinity of the BH-disk system (the geometry and location of the corona is still not well understood). Photons that pass into the corona can upscatter through inverse Compton scattering with high-energy electrons. The resultant spectrum is well-described by a power-law and a part of this radiation returns to the disk. The returning power-law spectrum can then reflect off the disk, modifying the power-law with the addition of fluorescent emission lines below 10~keV and a Compton hump normally peaked at 20-30~keV~\cite{Ross:2005dm, Garcia:2010iz}. The lines, as the emission travels to the observer, are broadened and skewed due to the strong gravity of the BH spacetime, and so the observed reflection spectrum is encoded with properties of the BH~\cite{Laor:1991nc,Bambi:2017khi}. A number of BHs have been studied using XRS, and in particular the spins of more than twenty stellar-mass and about forty supermassive BHs have been estimated~\cite{Bambi:2020jpe}.

In addition to measuring the BH spin, XRS can be used to test the \textit{Kerr hypothesis}~\cite{Lu:2002vm,Schee:2008fc,Johannsen:2012ng,Bambi:2012at,Cao:2017kdq,Bambi:2015kza},~i.e.~that all isolated, stationary, and axisymmetric astrophysical (uncharged) BHs are described by the Kerr metric. The Kerr hypothesis holds in general relativity, but may be violated in some modified gravity theories (see, e.g.,~\cite{Alexander:2009tp} and~\cite{Kleihaus:2011tg}) or with the introduction of new physics, such as large quantum effects~\cite{Giddings:2017jts} or exotic matter fields~\cite{Herdeiro:2014goa}. As there are a number of possible modifications to the Kerr metric that would violate the Kerr hypothesis, it has become useful to use parameterized non-Kerr metrics as a way to study possible departures rather than studying specific non-Kerr BH solutions. One popular parameterized metric that we have studied using XRS in numerous works (see, e.g.,~\cite{Tripathi:2018lhx,Tripathi:2019fms,Zhang:2019ldz,Zhang:2020qbx,Tripathi:2020dni,Tripathi:2020yts}) is the Johannsen metric~\cite{Johannsen:2015pca}.

In this work, we continue the study of the Konoplya-Rezolla-Zhidenko (KRZ) parameterized BH metric~\cite{Konoplya:2016jvv}, earlier done with the supermassive BHs in Seyfert-1 galaxies Ark~564 in Ref.~\cite{Nampalliwar:2019iti} (wherein we also analyze the effect of the deformation parameters on the reflection spectrum and present the implementation of the KRZ metric in the data analysis framework) and MCG--06--30--15 in Ref.~\cite{Abdikamalov:2021zwv} (hereafter Paper~I).
Here we instead study a stellar-mass BH, and we can thus probe a different curvature regime. For example, in models with higher curvature corrections to general relativity, lighter BHs can provide more stringent tests than heavier BHs because the curvature at the BH horizon scales as the inverse of the square of the BH mass~\cite{Yagi:2016jml}. For such a purpose, we consider a \textsl{NuSTAR} observation of the stellar-mass BH in the X-ray binary EXO~1846--031 during the outburst of 2019. Such an observation is particularly suitable for testing the Kerr hypothesis: the source was very bright, with a simple spectrum showing strong reflection features, and the quality of the \textsl{NuSTAR} data is very good. In a forthcoming paper (Swarnim Shashank et al., hereafter Paper~III), we will constrain the KRZ deformation parameters with \textsl{LIGO} and \textsl{Virgo} gravitational wave data, and we will thus compare the constraining power of XRS and gravitational waves.

The paper is organized as follows. In Section~\ref{s-krz}, we briefly describe the KRZ metric. In Section~\ref{s-red}, we summarize the observations to be studied and explain the data reduction performed. Section~\ref{s-ana} goes through the spectral analysis and results. Lastly, we discuss our results in Section~\ref{s-dis}.


\section{The KRZ parametrization}
\label{s-krz}

We begin with a review of the KRZ metric and its implementation in the \texttt{relxill\_nk} package~\cite{Bambi:2016sac,Abdikamalov:2019yrr,Abdikamalov:2020oci}. More details can be found in Refs.~\cite{Konoplya:2016jvv,Nampalliwar:2019iti,Ni:2016uik,Abdikamalov:2021zwv}. The KRZ metric uses a continued-fraction-based parametrization of stationary and axisymmetric spacetimes, with asymptotic flatness imposed \textit{a posteriori}. It enjoys several advantages over other parametrized non-Kerr BHs. In particular, a continued-fraction-based parametrization allows better convergence than the typical power-series-based parametrization. Moreover, the KRZ metric is not required to possess the Carter constant found in Kerr, and typical for non-Kerr, metrics, allowing for more generic deviations.

In Boyer-Lindquist-like coordinates, the line element of the KRZ metric is written as~\cite{Konoplya:2016jvv}
\bwide
\be\label{eq:metric}
ds^2 &=& - \frac{N^2 - W^2 \sin^2\theta}{K^2} \, dt^2 - 2 W r \sin^2\theta \, dt \, d\phi
+ K^2 r^2 \sin^2\theta \, d\phi^2 
+ \frac{\Sigma \, B^2}{N^2} \, dr^2 + \Sigma \, r^2 \, d\theta^2 \, ,
\ee
\ewide
where $\Sigma = 1+a_*^2\cos^2{\theta}/r^2$, $a_* = J/M^2$ is the dimensionless spin parameter, and $N, W, K$ and $B$ are functions quantifying deviations away from Kerr. They depend on $r$ and $\theta$ and are written to facilitate asymptotic matching with the Kerr metric (e.g., through parametrized post-Newtonian constraints). For instance, $W$ is given as
\be
    W = \frac{1}{\Sigma}\sum_{i=0}^{\infty} \left \{ w_{i0} (1-x)^2 + \widetilde{W}_i (x) (1-x)^3 \right \}y^i, 
\ee
where $x = 1-r_0/r$, $r_0$ is the radius of the event horizon, $y = \cos\theta$, and 
\be
    \widetilde{W}_i (x) = \frac{w_{i1}}{1+\frac{w_{i2}x}{1+\frac{w_{i3}x}{1+\cdots}}}.
\ee

In its implementation in \texttt{relxill\_nk}, we choose to analyze the deviation parameters appearing at the leading order, following the choice made in Ref.~\cite{Ni:2016uik}. The deviation functions are written as follows~\cite{Nampalliwar:2019iti}
\bwide
\be\label{eq:deffunc}
N^2 &=& \left(1 - \frac{r_0}{r}\right) \left (1 - \frac{\epsilon_0 r_0}{r} 
+ \left(k_{00} - \epsilon_0\right)\frac{r_0^2}{r^2} + \frac{\delta_1 r^3_0}{r^3}\right ) 
+ \left( \frac{a_{20} r^3_0}{r^3}
+ \frac{a_{21} r^4_0}{r^4} + \frac{k_{21} r^3_0}{r^3 \left ( 1+ \frac{k_{22}(1-\frac{r_0}{r}) }{1+k_{23}(1-\frac{r_0}{r})}\right)}\right) \cos^2\theta  \, , \\
B &=& 1 + \frac{\delta_4 r^2_0}{r^2} + \frac{\delta_5 r^2_0}{r^2} \cos^2\theta \, , \qquad
W = \frac{1}{\Sigma} \left(\frac{w_{00} r^2_0}{r^2} + \frac{\delta_2 r^3_0}{r^3}
+ \frac{\delta_3 r^3_0}{r^3} \cos^2\theta \right) \, , \qquad \\
K^2 &=& 1 + \frac{a_* W}{r} + \frac{1}{\Sigma} \left( \frac{k_{00} r^2_0}{r^2} 
+ \left ( \frac{k_{20} r^2_0}{r^2} + \frac{k_{21} r^3_0}{r^3 \left ( 1+ \frac{k_{22}(1-\frac{r_0}{r})}{1+k_{23}(1-\frac{r_0}{r})}\right ) } \right ) \cos^2\theta \right)\, .
\ee
\ewide
where
\begin{align}
\begin{aligned}
r_0 = 1 + \sqrt{1 - a_*^2} \, , \qquad \epsilon_0 = \frac{2 - r_0}{r_0} \, ,  \\ 
a_{20} = \frac{2 a_*^2}{r^3_0} \, , \qquad a_{21} = - \frac{a_*^4}{r^4_0} + \delta_6 \, ,\\
k_{00} = k_{22} = k_{23} = \frac{a_*^2}{r_0^2} \, , \qquad  k_{20} = 0 \, ,  \\
k_{21} = \frac{a_*^4}{r_0^4} - \frac{2 a_*^2}{r^3_0} - \delta_6 \, , \qquad w_{00} = \frac{2 a_*}{r^2_0} \, .
\end{aligned}
\end{align}
Deviations from the Kerr metric are parametrized by the six deformation parameters $\{ \delta_i \}$ ($i = 1, 2, ... 6$). $\delta_1$ relates to the deformations of $g_{tt}$, $\delta_2$ and $\delta_3$ to deformations related to the BH rotation, $\delta_4$ and $\delta_5$ to deformations of $g_{rr}$, and $\delta_6$ affects the shape of the event horizon. Only $\delta_1$ and $\delta_2$ affect the size of the innermost stable circular orbit or ISCO (see, e.g., Ref.~\cite{Nampalliwar:2019iti} for the ISCO contours). Similarly, the reflection spectrum is  sensitive primarily to $\delta_1$ and $\delta_2$, other parameters affecting it at a moderate to negligible level.

The \texttt{relxill\_nk} suite of models vary one non-Kerr parameter at a time. The range for each parameter is chosen to ensure no pathologies (e.g., closed timelike curves, singularities outside the event horizon) appear in the spacetime outside the event horizon. The allowed range of each $\{ \delta_i \}$ ($i = 1, 2, ... 5$) is as follows~\cite{Nampalliwar:2019iti}
\begin{gather}
	\delta_1 > \frac{4r_0 - 3r_0^2 - a_*^2}{r_0^2}, \\\nonumber
	\delta_2, \delta_3 \left\{\begin{array}{l}
		> \\
		<\\
		\end{array} -\frac{4}{a_*^3}(1-\sqrt{1-a_*^2}) \quad \begin{array}{l}
		 \rm{if} \; a_* > 0\\ 
		 \rm{if} \; a_* < 0,
		\end{array} \right. \\\nonumber
	\delta_4, \delta_5 > -1.
\end{gather}
The same for on $\delta_6$ cannot be expressed analytically, but is easily calculated numerically. In \texttt{relxill\_nk}, we use the above bounds or $\pm 5$ (except in the case of $\delta_1$, where we use $+4$ as the upper bound), whichever is stronger, for each deformation parameter. In the following sections, we discuss astrophysical bounds on each of these parameters obtained by analyzing the X-ray data from an astrophysical source.


\section{Observations and data reduction \label{s-red}}

EXO~1846--031 is a low mass X-ray binary (LMXB) discovered by the \textsl{EXOSAT} mission on April 3rd, 1985 \cite{parmar1985exo}. It has been quiescent for over two decades since at least 1994 \cite{1994EXO}. On July 23rd, 2019, \textsl{MAXI} detected a hard X-ray transient consistent with the location of EXO~1846--031, indicating a renewed activity of the BH candidate in EXO~1846--031 \cite{negoro2019maxi}. Soon it was located by \textsl{Swift}/XRT \cite{mereminskiy2019localization} and detected in radio bands by VLA \cite{miller2019vla} and MeerKAT \cite{williams2019meerkat}.

On August 3, 2019, \textsl{NuSTAR} observed EXO~1846--031 in an TOO observation for 22.2~ks with the FPMA and FPMB detectors, under ObsID 90501334002. This observation was first analyzed in Ref.~\cite{Draghis:2020ukh}, where the authors assumed the Kerr metric and measured the BH spin. Our analysis follows that of Ref.~\cite{Draghis:2020ukh}.

The raw data are downloaded from the HEASARC website and processed into cleaned event files using the routines in HEASOFT v6.26.1 through the NuSTARDAS pipeline v1.8.0 and CALDB v20190812. The source region is extracted with a radius of 180 arcsec in the two sensors and the extraction of the background region is the same. In this process, we also apply the ftool \texttt{grppha} to group the data to make sure each energy bin contains more than 30 counts. Previous studies on this observation of EXO~1846--031 reveal that the system has a high spin ($a_*\sim0.997$), the source is observed in a high-luminosity hard state, where the inner radius $R_\mathrm{in}$ is close to $R_\mathrm{ISCO}$, and the relativistic reflection features are strong \cite{Draghis:2020ukh,Wang:2020jcd}. All these ingredients make this observation/source ideal for testing the Kerr hypothesis.


\begin{figure*}
\includegraphics[width=0.65\textwidth]{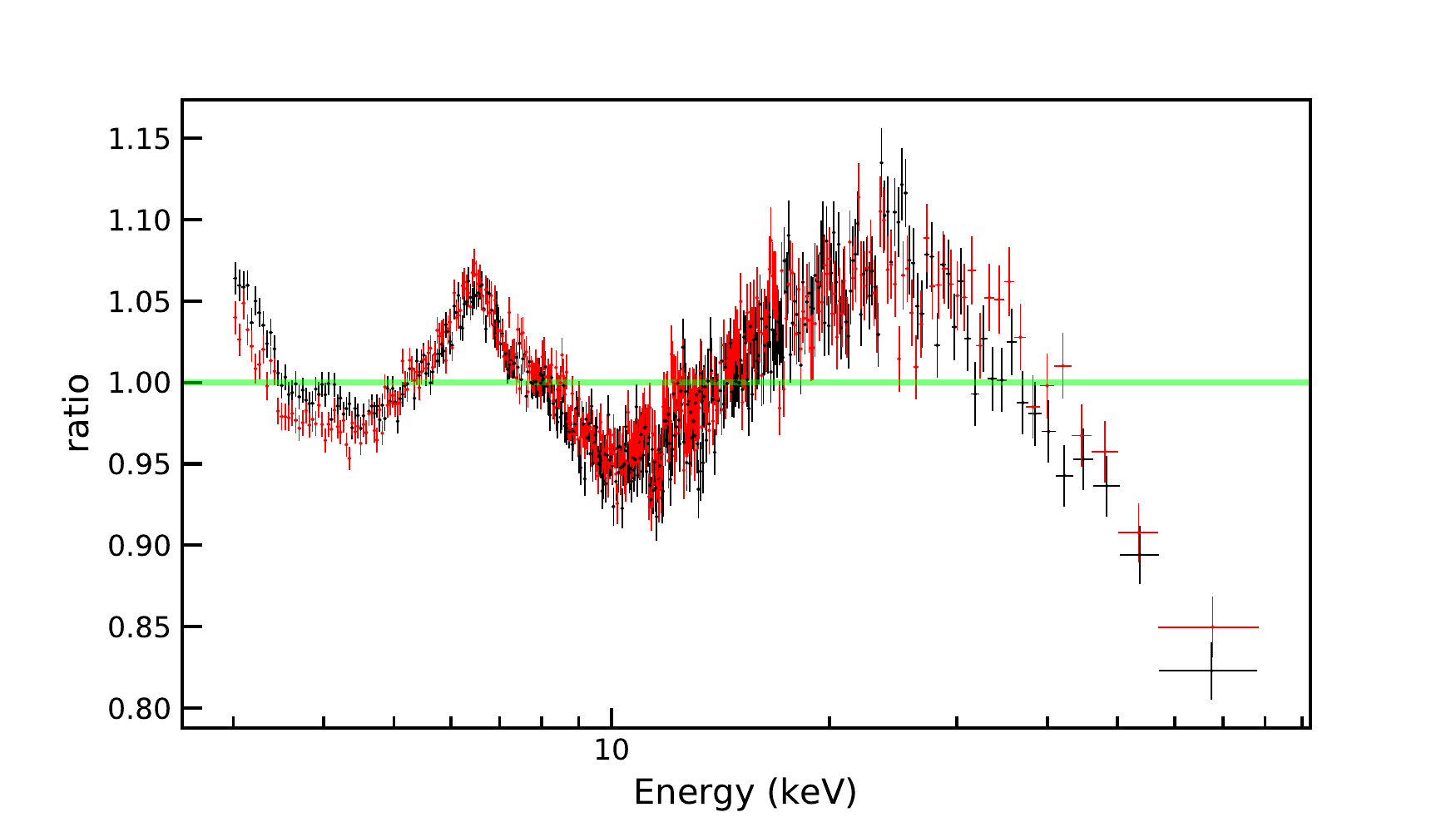}
\vspace{-0.3cm}
    \caption{Data to best-fit model ratio for an absorbed power-law. We clearly see a broad iron line at 5-8~keV and a Compton hump peaked at 20-30~keV indicating the presence of prominent reflection features in the spectrum. Black and red crosses are for FPMA and FPMB data, respectively. \label{fig:ratio-cutoffpl}}
\end{figure*}

\section{Spectral analysis and constraints \label{s-ana}}

We employ Xspec v12.10.1s \cite{1996ASPC..101...17A}, WILMS abundance \cite{Wilms:2000ez} and VERN cross-section \cite{Verner:1996th} in our spectral analysis. We first fit the data with an absorbed power-law model \texttt{tbabs*cutoffpl}. \texttt{tbabs} accounts for the interstellar absorption and has the column density $N_\mathrm{H}$ as the only parameter; \texttt{cutoffpl} describes a power-law continuum with photon index $\Gamma$, high energy cut-off $E_\mathrm{cut}$, and a normalization parameter. The data to best-fit model ratio is shown in Fig.~\ref{fig:ratio-cutoffpl}. A thermal component is clearly seen in the 3-4 keV soft band. A blurred Fe K$\alpha$ line feature at 5-8 keV together with a Compton hump at 20-30 keV indicate a relativistic reflection component.

To better account for the residuals, we add \texttt{diskbb}~\cite{Mitsuda:1984nv} to fit the thermal spectrum from the disk above 3 keV and we use the relativistic reflection model \texttt{relxillion\_nk}~\cite{Abdikamalov:2021rty} to fit the reflection features. \texttt{diskbb} is a multi-temperature blackbody  model with the temperature at inner disk radius ($kT_\mathrm{in}$) and the normalization of the component. \texttt{relxillion\_nk} is a new flavor in the model package \texttt{relxill\_nk} implemented with a radial disk ionization profile \citep{Abdikamalov:2021rty}. In this work, we apply the KRZ metric as the background metric.

Following the report of \citet{Madsen:2020gzx} that a tear in the \textsl{NuSTAR} thermal blanket caused an increased flux in FPMA, we ignore the 3-7 keV band of FPMA and leave a constant offset between the two detectors ($C_\mathrm{FPMA}=1$) when doing the original fit. Then we apply the multiplicative table \texttt{nuMLIv1.mod} provided by \textsl{NuSTAR} SOC to account for the discrepancy of the fluxes between FPMA and FPMB. We set the MLI covering fraction of FPMB to 1 and allow the covering fraction of FPMA to vary. After this, we fix the constant offset of FPMB and notice the 3-7 keV band of FPMA to continue our fitting. In Xspec, the full model reads as:

\vspace{0.2cm}
    
\noindent \texttt{mtable\{nuMLIv1.mod\}*tbabs*(diskbb+relxillion\_nk)}.

\vspace{0.2cm}

The relativistic reflection model \texttt{relxillion\_nk} returns both the reflection component and the Comptonized power-law component from the corona when the reflection fraction ($R_{\rm f}$) is set to be positive \cite{Dauser:2016yuj}. The parameters describing the power-law spectrum from the corona are $\Gamma$ and $E_\mathrm{cut}$, as in \texttt{cutoffpl}. The reflection model assumes a broken power-law emissivity profile $\epsilon(r)$: $\epsilon\propto r^{-q_\mathrm{in}}$ for $r<R_\mathrm{br}$ and $\epsilon\propto r^{-q_\mathrm{out}}$ for $r>R_\mathrm{br}$, where $q_\mathrm{in}$ and $q_\mathrm{out}$ are inner and outer emissivity indices and $R_\mathrm{br}$ is the breaking radius on the disk. This is a phenomenological assumption when the geometry of the corona is unknown. First, we allow both $q_\mathrm{in}$ and $q_\mathrm{out}$ to vary in the fit. However, we always get an almost flat outer emissivity index ($q_\mathrm{out}\sim0$) and therefore we fix $q_\mathrm{out}=0.1$ to save computational power. In \texttt{relxillion\_nk}, the ionization of the disk is
\begin{equation}
    \xi(r)=\xi_0\qty(\frac{R_\mathrm{in}}{r})^{\alpha_\xi},
\end{equation}
where $\xi_0$ in the ionization parameter at the inner edge of the accretion disk and $\alpha_\xi$ provides the radial profile.

If we use the basic version of \texttt{relxill\_nk}, in which the ionization parameter is assumed to be constant over the disk, we find that we need to add a Gaussian absorption line around 7~keV to model a feature in the spectrum. Moreover, we find that the reflection fraction $R_{\rm f}$ cannot be constrained. Both findings are in agreement with the study reported in Ref.~\cite{Draghis:2020ukh}. If we use \texttt{relxillion\_nk}, we do not need any absorption line and the reflection fraction can be constrained to a value close to 1, which is the value expected for a corona with isotropic emission. We also note that the fit obtained with \texttt{relxillion\_nk} has a lower $\chi^2$ ($\Delta\chi^2\sim-20$) than the fit with \texttt{relxill\_nk+gaussian}. We thus use the model with non-trivial ionization gradient.

\begin{figure*}
\includegraphics[width=0.48\textwidth]{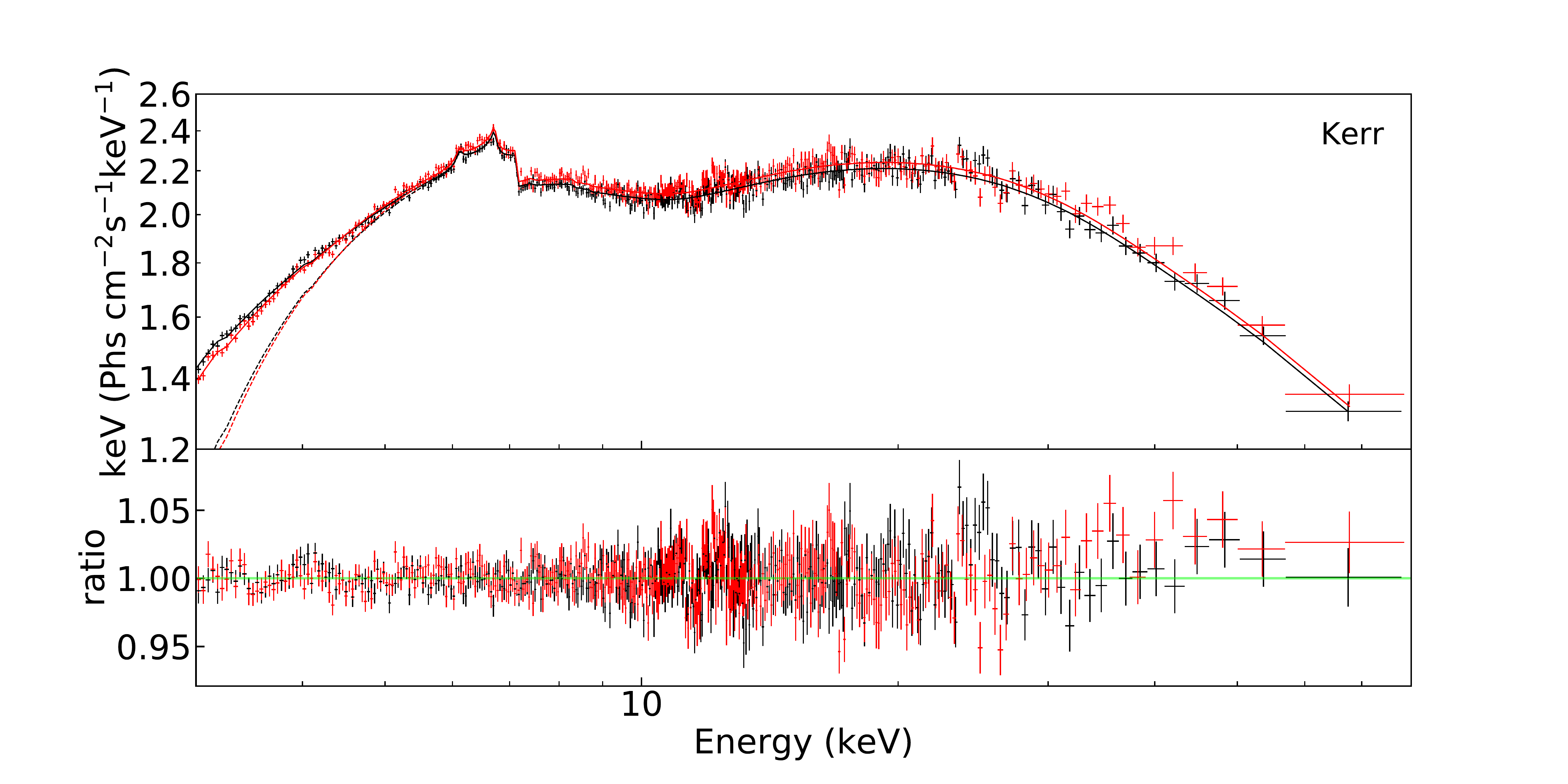}
\includegraphics[width=0.48\textwidth]{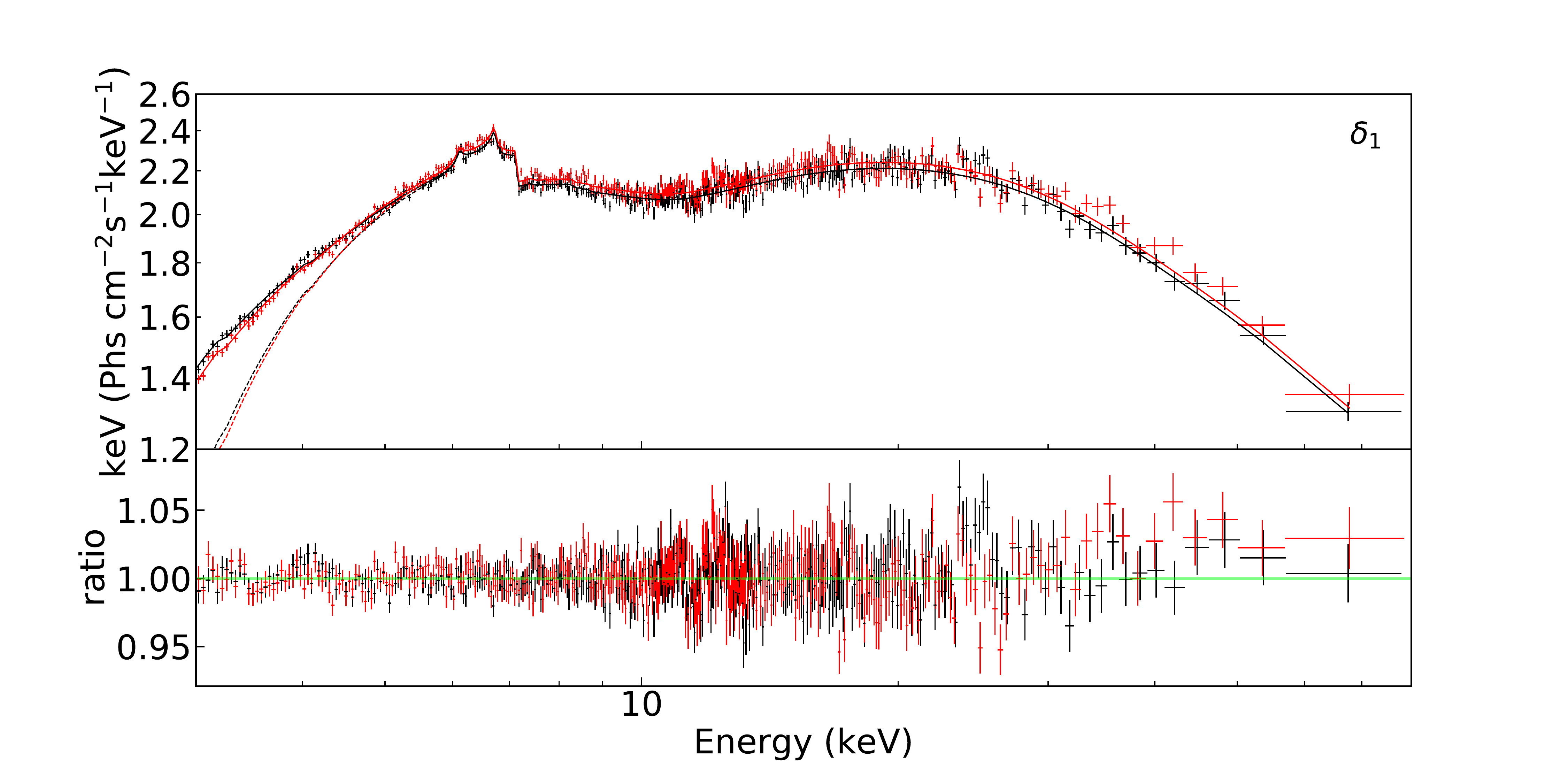}
\includegraphics[width=0.48\textwidth]{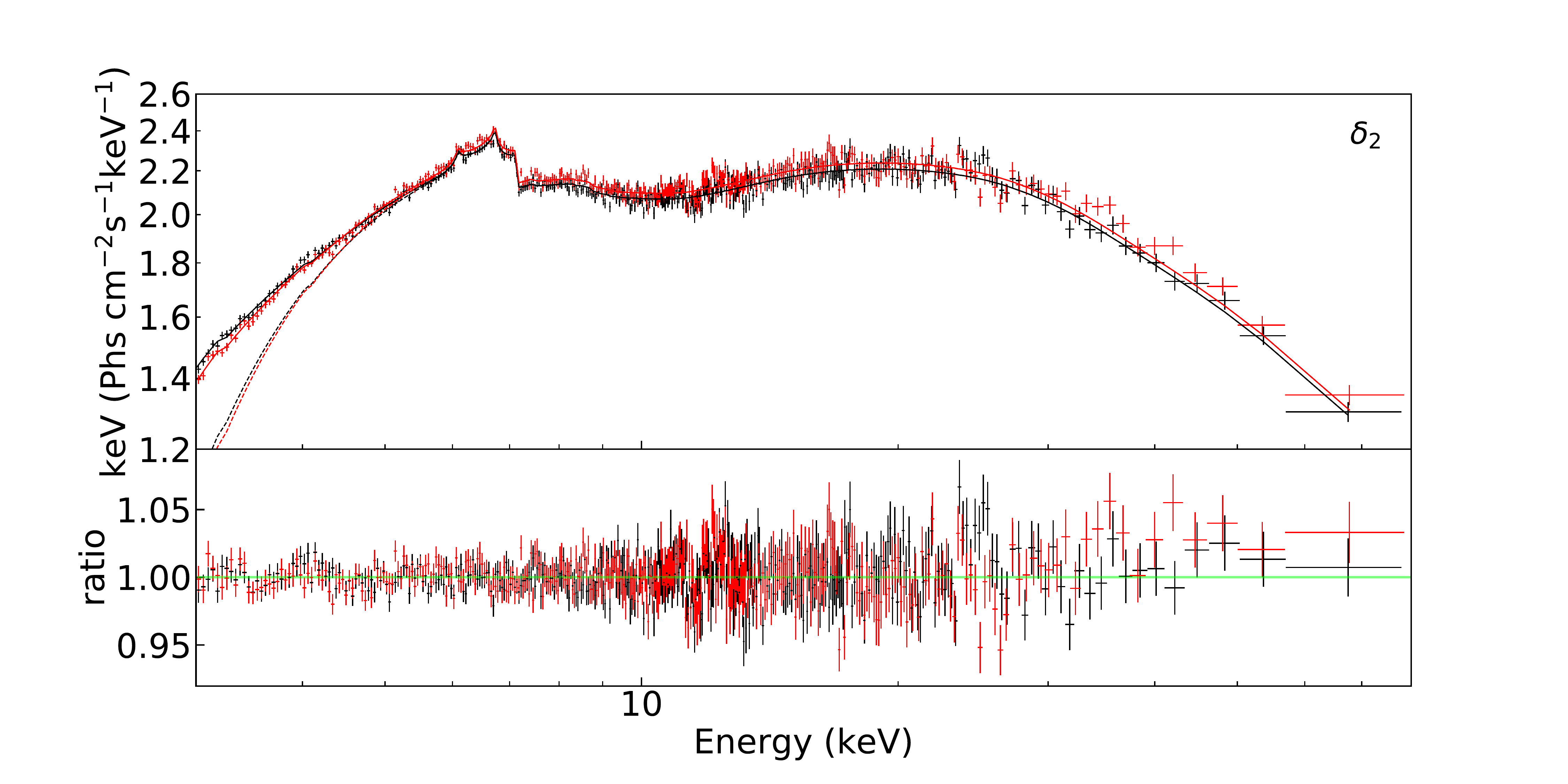}
\includegraphics[width=0.48\textwidth]{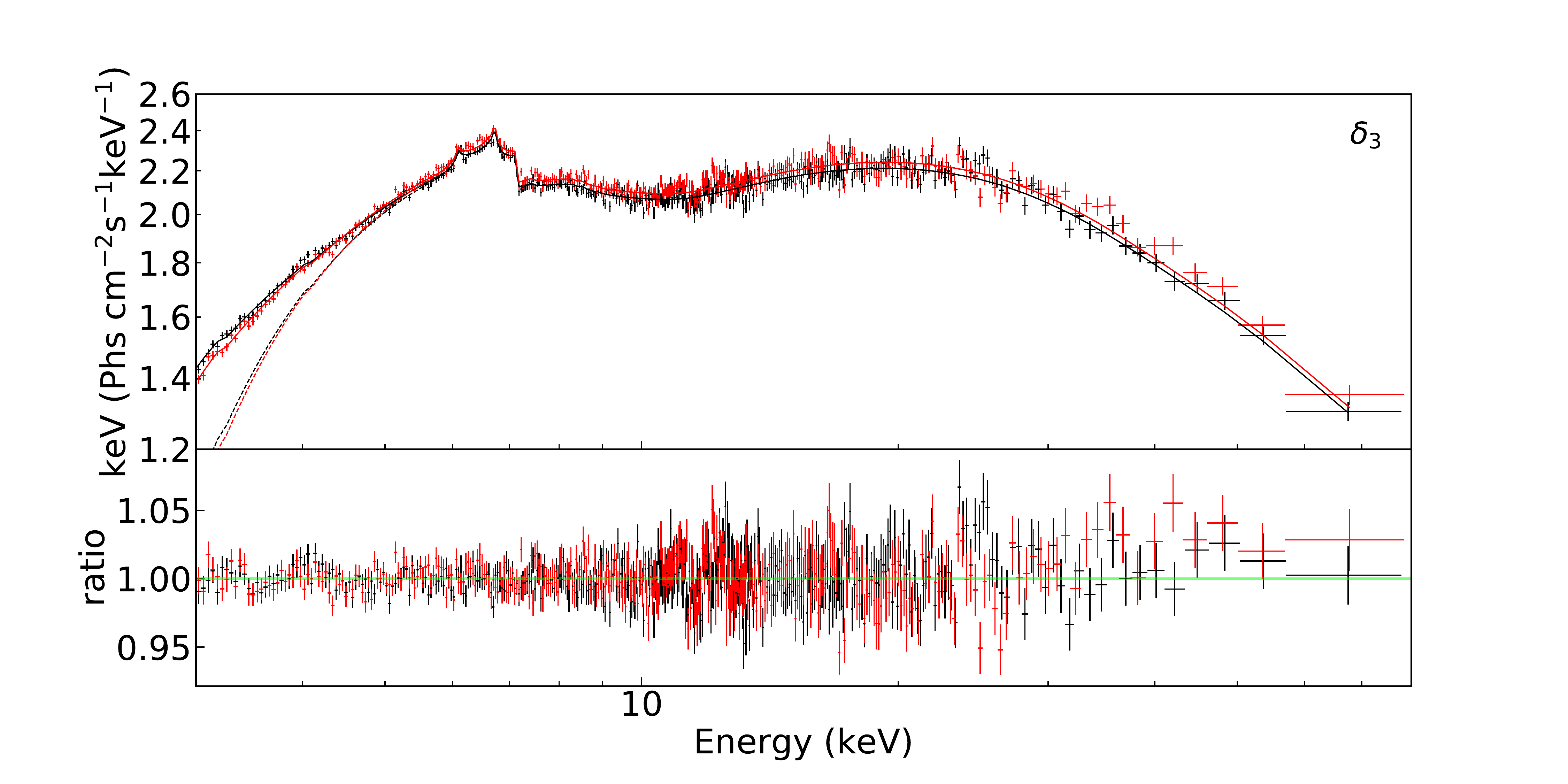}
\includegraphics[width=0.48\textwidth]{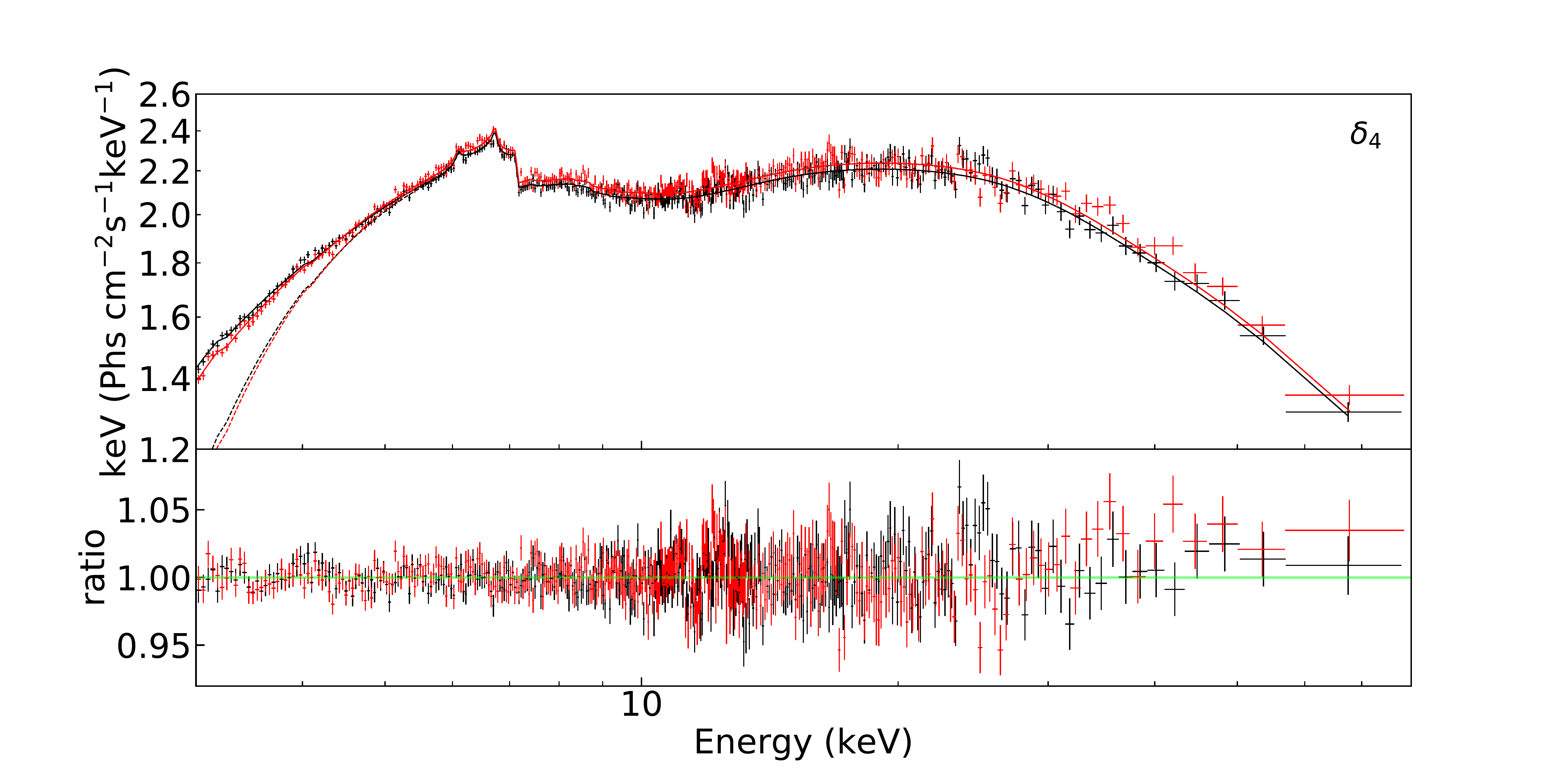}
\includegraphics[width=0.48\textwidth]{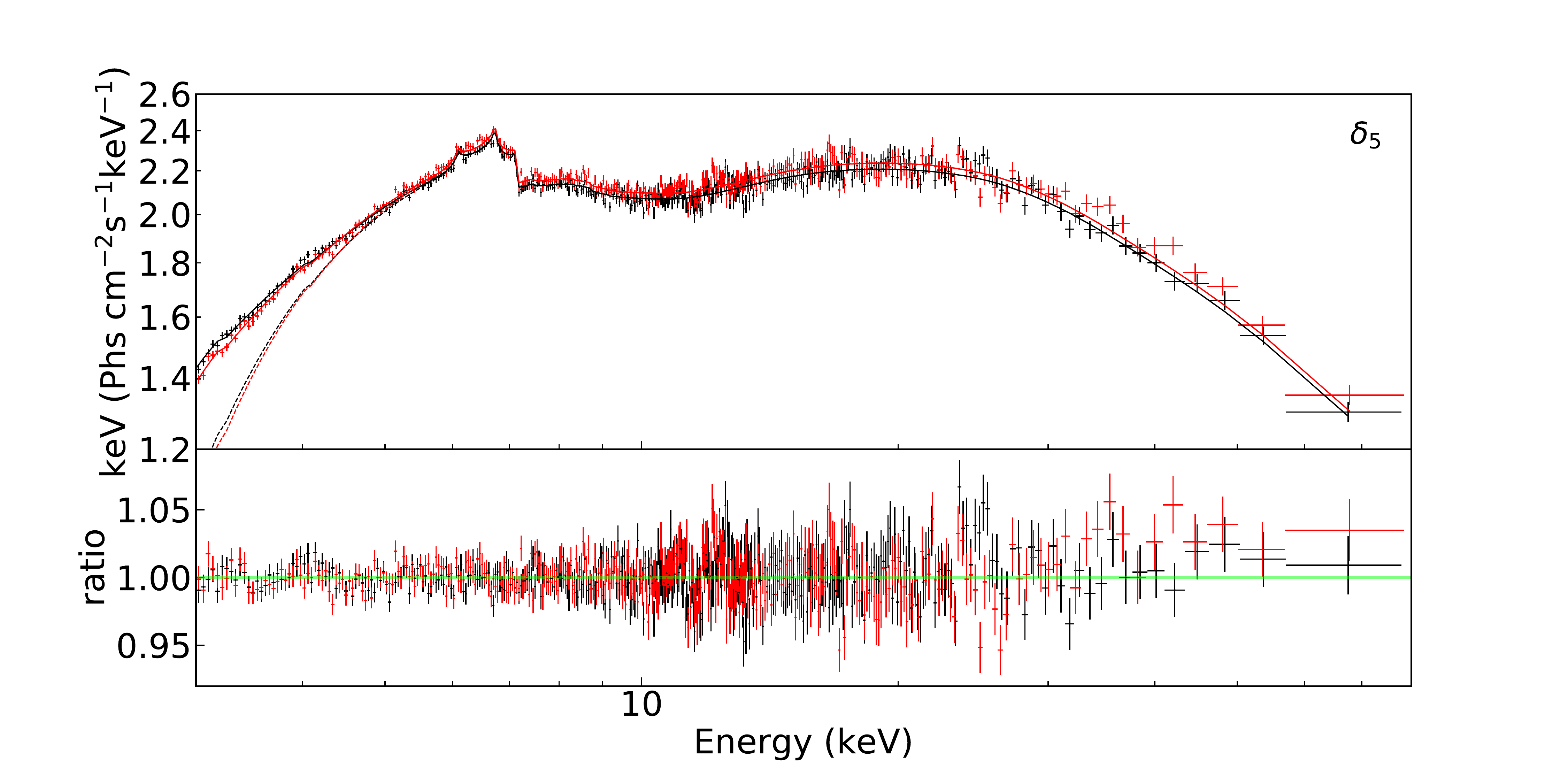}
\vspace{-0.3cm}
    \caption{Best-fit models (upper quadrants) and data to best-fit model ratios (lower quadrants) for the Kerr BH scenario (top-left panel) and KRZ BH models with different deformation parameters. Black and red crosses are for FPMA and FPMB data, respectively.  \label{fig:eeuf-ratio}}
\end{figure*}

\begin{figure*}
\includegraphics[width=0.48\textwidth,trim={1.0cm 1.5cm 2.0cm 2.0cm},clip]{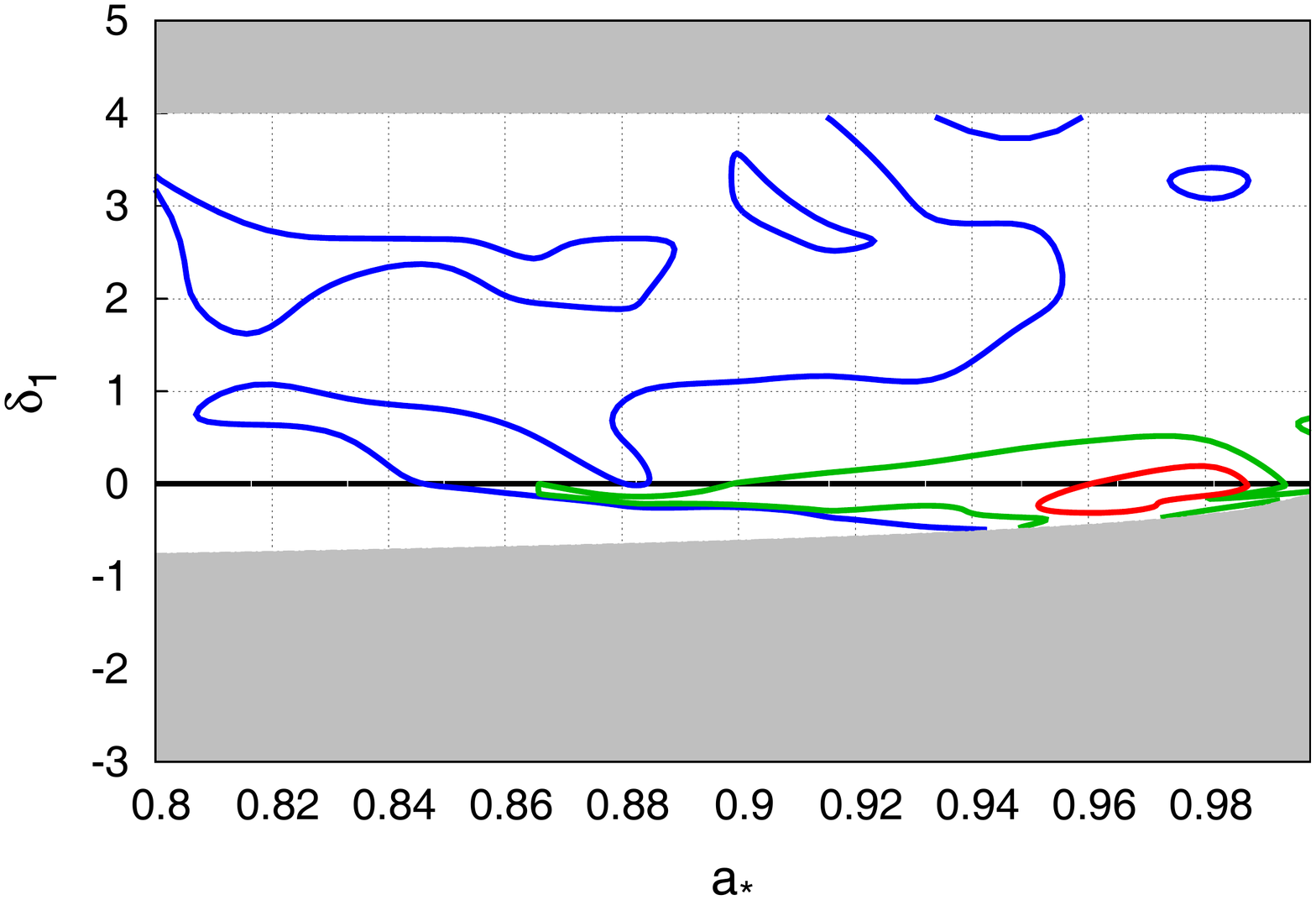}
\includegraphics[width=0.48\textwidth,trim={1.0cm 1.5cm 2.0cm 2.0cm},clip]{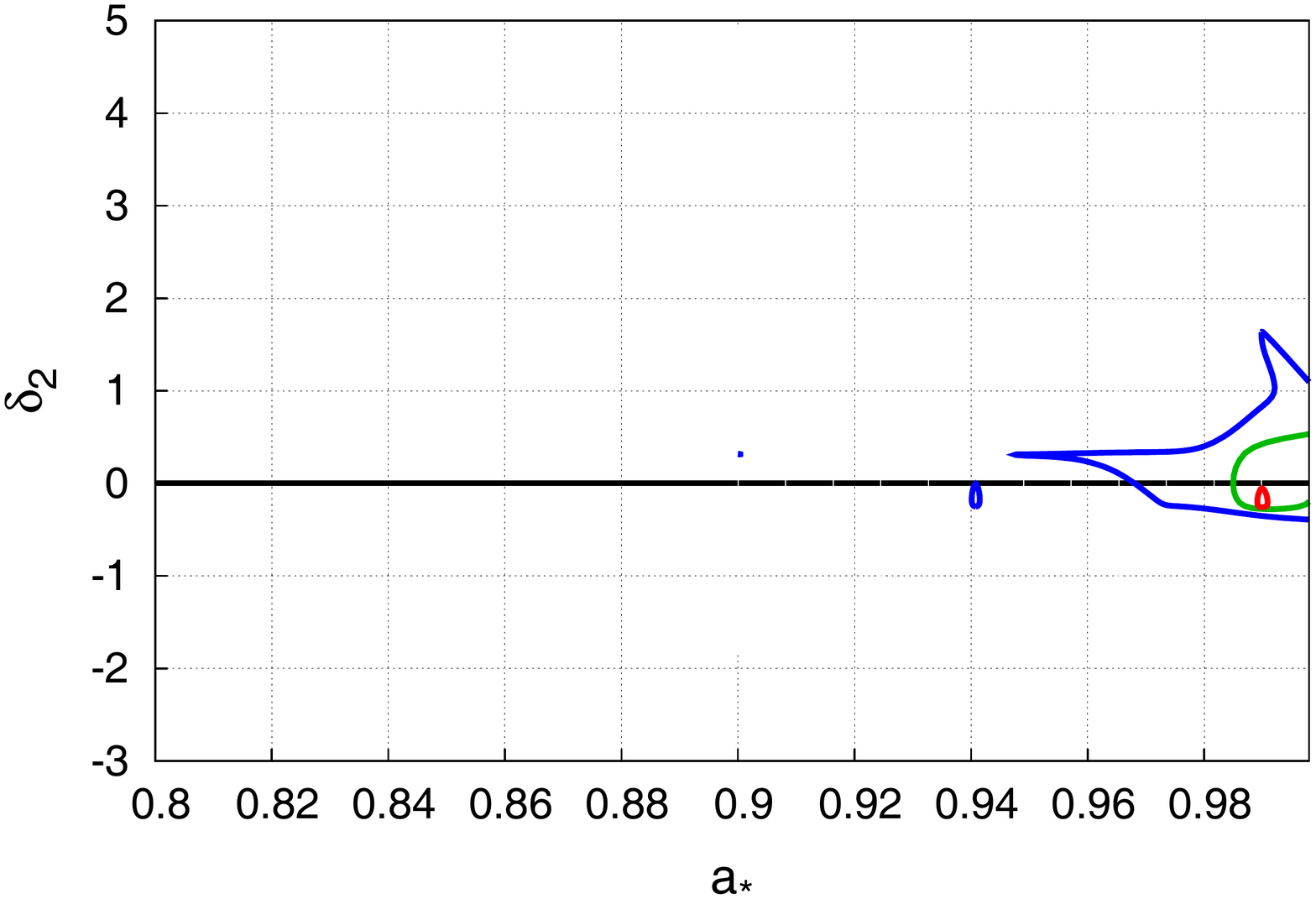}
\includegraphics[width=0.48\textwidth,trim={1.0cm 1.5cm 2.0cm 2.0cm},clip]{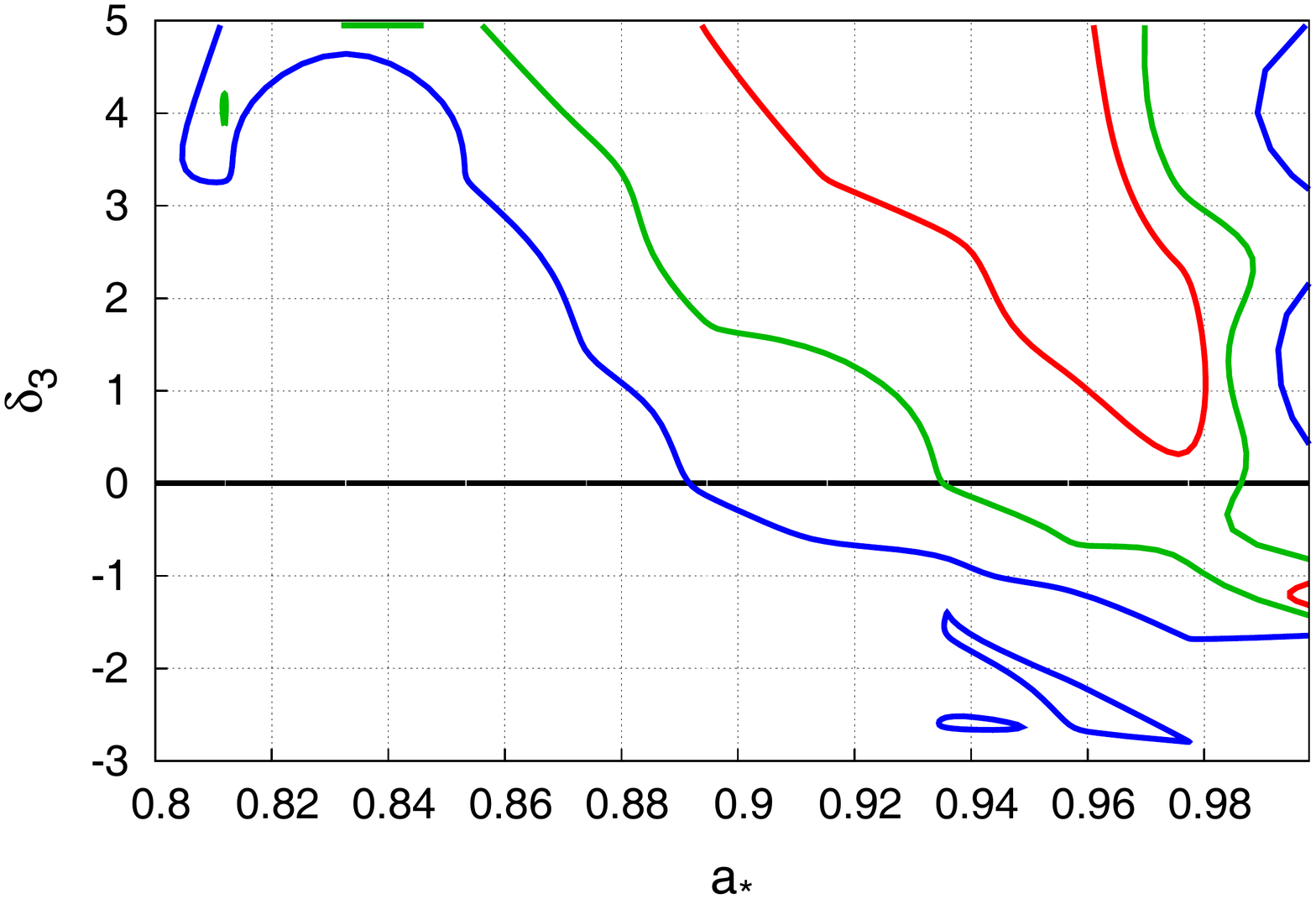}
\includegraphics[width=0.48\textwidth,trim={1.0cm 1.5cm 2.0cm 2.0cm},clip]{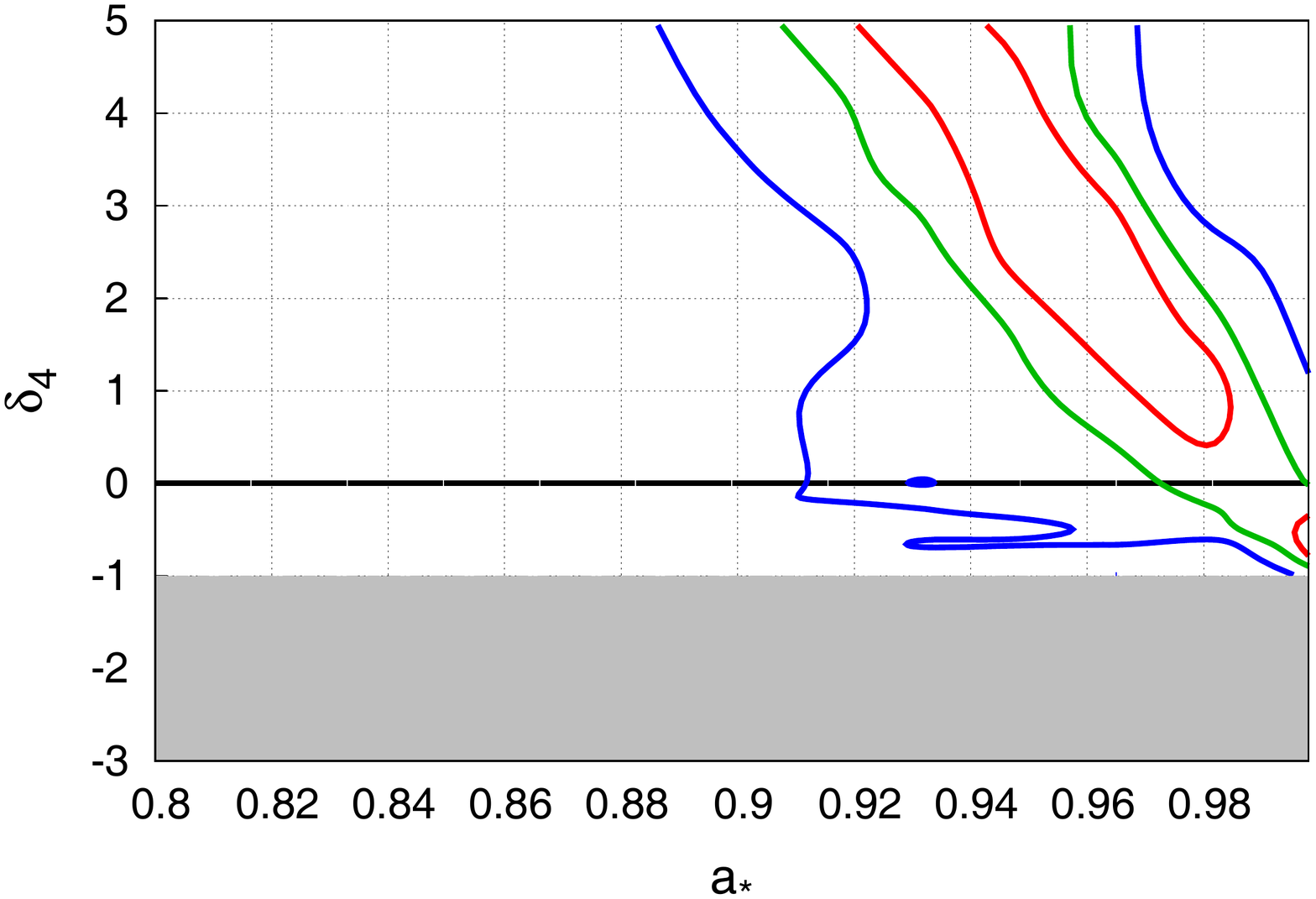}
\includegraphics[width=0.48\textwidth,trim={1.0cm 1.5cm 2.0cm 2.0cm},clip]{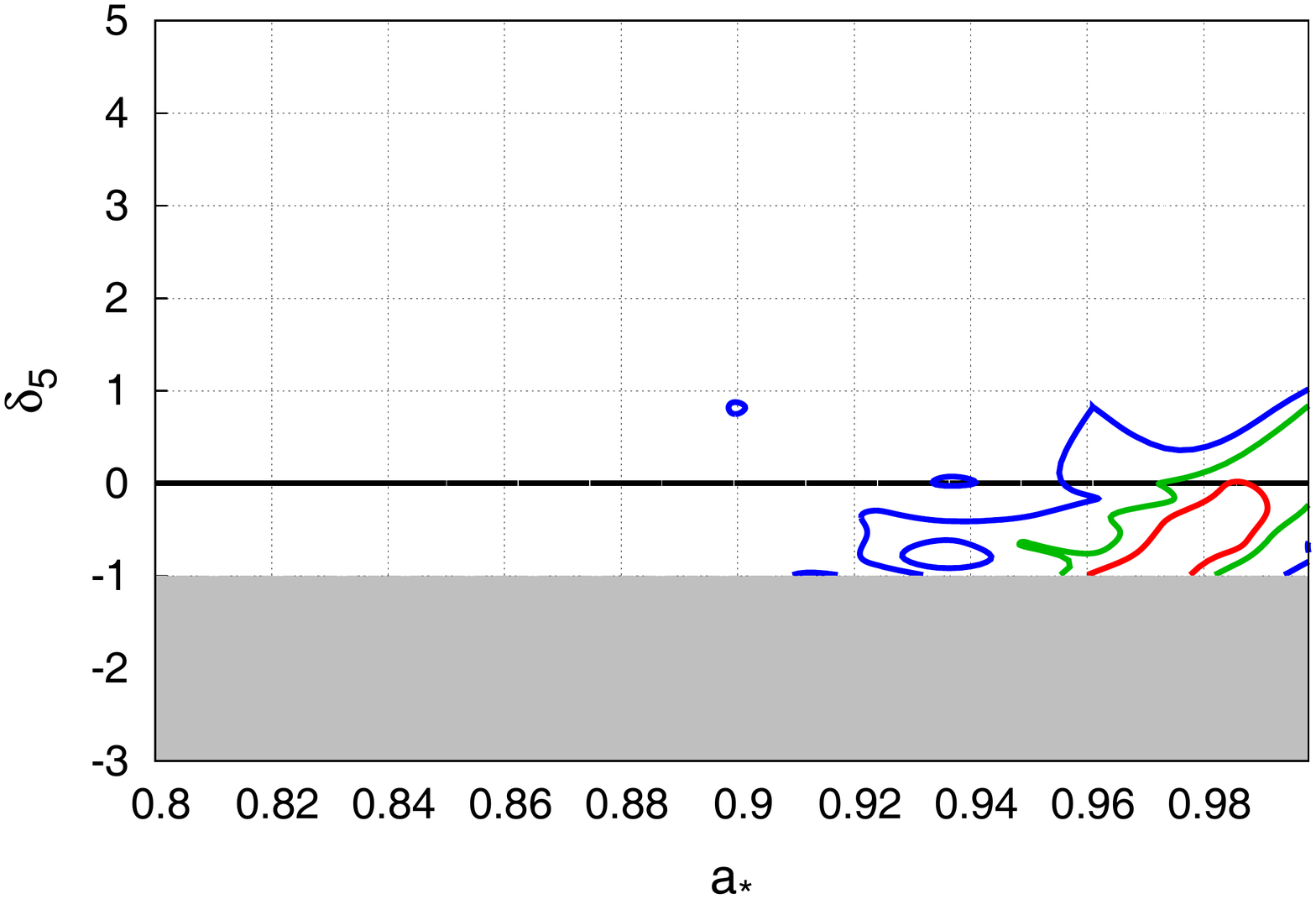}
\vspace{-0.4cm}
    \caption{Constraints on the BH spin parameter $a_*$ and on the KRZ deformation parameters from the analysis of the reflection features of the BH in EXO~1846--031 with the model \texttt{relxillion\_nk}. The red, green, and blue curses correspond, respectively, to the 68\%, 90\%, and 99\% confidence level contours for two relevant parameters. 
    \label{fig:contours}}
\end{figure*}

Tab.~\ref{tab:fit-relxillion_nk} shows the results of our fit with the Kerr metric and with a possible non-vanishing deformation parameter \{$\delta_i$\} ($i=1,2,...5$). The current version of \texttt{relxill\_nk} cannot have two free deformation parameters at the same time, so we analyze the data assuming that only one of the deformation parameters may be non-vanishing and we set the values of all other deformation parameters to zero. Unlike in Paper~I on MCG--06--30--15, we are not able to to get any result for the deformation parameter $\delta_6$. We have not been able to fix the problem in the model and we leave the analysis of the case with free $\delta_6$ to a possible future work. Fig.~\ref{fig:eeuf-ratio} shows the best-fit models (upper quadrants) and the data to best-fit model ratios (lower quadrants) of our six fits. Last, Fig.~\ref{fig:contours} shows the constraints on the BH spin parameter $a_*$ and the five KRZ deformation parameters after marginalizing over all other model parameters. The red, green, and blue curves indicate, respectively, the 68\%, 90\%, and 95\% confidence level limits for two relevant parameters. The discussion of our results is postponed to the next section.

\begin{table*}
    \centering
    \renewcommand\arraystretch{1.3}{
    \begin{tabular}{lcccccc}
        \hline\hline
        & Kerr & $\delta_1$ & $\delta_2$ & $\delta_3$ & $\delta_4$ & $\delta_5$\\
        \hline
        \texttt{nuMLI.mod} &&&&&& \\
        $cf_A$ & $0.882_{-0.014}^{+0.014}$ & $0.882_{-0.015}^{+0.015}$ & $0.882_{-0.15}^{+0.15}$ & $0.882_{-0.015}^{+0.015}$ & $0.882_{-0.015}^{+0.015}$ & $0.882_{-0.015}^{+0.015}$ \\
        \hline
        \texttt{tbabs} &&&&&& \\
        $N_{\text{H}}$/$10^{22}$ cm$^{-2}$ & $7.7_{-0.9}^{+0.9}$ & $7.6_{-0.8}^{+0.8}$ & $7.9_{-0.8}^{+0.9}$ & $7.6_{-0.8}^{+0.8}$ & $8.0_{-0.9}^{+0.9}$ & $8.0_{-0.8}^{+0.9}$\\
        \hline
        \texttt{diskbb} &&&&&& \\
        $kT_{\text{in}}$ [keV] & $0.43_{-0.03}^{+0.02}$ & $0.43_{-0.03}^{+0.02}$ & $0.412_{-0.014}^{+0.020}$ & $0.43_{-0.03}^{+0.02}$ & $0.41_{-0.03}^{+0.02}$ & $0.41_{-0.03}^{+0.02}$\\
        norm [$10^{4}$] & $1.9_{-0.6}^{+1.1}$ & $1.9_{-0.6}^{+0.9}$ & $2.6_{-0.7}^{+1.5}$ & $1.9^{+1.5}_{-0.7}$ & $2.7_{-0.7}^{+1.3}$ & $2.7_{-0.8}^{+1.4}$\\
        \hline
        \texttt{relxillion\_nk} &&&&&& \\
        $q_{\rm in}$ & $7.6_{-0.6}^{+2.2}$ & $7.9_{-0.8}^{\rm +(P)}$ & $9_{-4}^{\rm +(P)}$ & $9.7_{-1.2}^{+0.1}$ & $9.0_{-1.5}^{\rm +(P)}$ & $8.7_{-0.5}^{\rm +(P)}$\\
        $q_{\rm out}$ & $0.1^*$ & $0.1^*$ & $0.1^*$ & $0.1^*$ & $0.1^*$ & $0.1^*$\\
        $R_{\text{br}}$ [$r_{\text{g}}$] & $12_{-2}^{+7}$ & $11_{-2}^{+10}$ & $8_{-1}^{+7}$ & $8_{-1}^{+2}$ & $8_{-1}^{+5}$ & $8_{-2}^{+5}$\\
        $a_*$ & $0.980_{-0.042}^{+0.010}$ & $0.973_{-0.009}^{+0.016}$ & $0.997_{-0.008}$ & $0.974_{-0.071}^{+0.017}$ & $0.982_{-0.041}^{+0.012}$ & $0.988_{-0.025}^{+0.005}$\\
        $\delta_i$ & $0^*$ & $-0.12_{-0.23}^{+0.30}$ & $-0.39_{-0.04}^{+0.08}$ & $1.9_{-1.9}^{\rm +(P)}$ & $0.8_{-1.3}^{+1.3}$ & $-0.1_{-0.4}^{+2.7}$\\
        $\iota$ [deg] & $64_{-4}^{+1}$ & $63.2_{-2.4}^{+2.0}$ & $71_{-4}^{+3}$ & $69_{-4}^{+2}$ & $71_{-2}^{+3}$ & $71_{-4}^{+2}$\\
        $\Gamma$ & $1.78_{-0.03}^{+0.05}$ & $1.78_{-0.03}^{+0.07}$ & $1.85_{-0.04}^{+0.03}$ & $1.78_{-0.03}^{+0.15}$ & $1.87_{-0.04}^{+0.07}$ & $1.87_{-0.04}^{+0.06}$\\
        $\log\xi_0$ & $4.7_{-0.5}$ & $4.7_{-0.3}$ & $4.5_{-0.3}^{+0.2}$ & $4.7_{-0.4}$ & $4.4_{-0.3}^{\rm +(P)}$ & $4.4_{-0.3}^{\rm +(P)}$\\
        $\alpha_\xi$ & $1.1_{-0.5}^{+0.7}$ & $1.2_{-0.4}^{+0.4}$ & $1.8_{-0.1}^{+0.4}$ & $1.9_{-0.3}^{+1.0}$ & $1.6_{-0.7}^{+0.4}$ & $1.5_{-0.4}^{+0.3}$\\
        $A_{\text{Fe}}$ & $8_{-2}^{\rm +(P)}$ & $8.0_{-4.2}^{+1.6}$ & $3.7_{-1.0}^{+2.6}$ & $7.0_{-5.1}^{+2.3}$ & $3.3_{-1.7}^{+5.2}$ & $3.0_{-1.6}^{+5.6}$\\
        $E_{\text{cut}}$ [keV] & $142_{-14}^{+21}$ & $141_{-7}^{+19}$ & $148_{-9}^{+12}$ & $140_{-15}^{+29}$ & $148_{-10}^{+26}$ & $149_{-21}^{+15}$\\
        $R_{\rm f}$ & $1.01_{-0.16}^{+0.26}$ & $1.03_{-0.18}^{+0.24}$ & $1.2_{-0.2}^{+0.5}$ & $1.05_{-0.17}^{+0.26}$ & $1.2_{-0.2}^{+0.5}$ & $1.2_{-0.3}^{+0.5}$\\
        norm [$10^{-2}$] & $0.72_{-0.11}^{+0.10}$ & $0.71_{-0.09}^{+0.09}$ & $0.73_{-0.14}^{+0.17}$ & $0.71_{-0.10}^{+0.10}$ & $0.74_{-0.12}^{+0.13}$ & $0.74_{-0.15}^{+0.17}$\\
        \hline
        $\chi^2/{\rm dof}$ & \hspace{0.2cm} $2717.14/2597$ \hspace{0.2cm} & \hspace{0.2cm} $2716.86/2596$ \hspace{0.2cm} & \hspace{0.2cm} $2715.98/2596$ \hspace{0.2cm} & \hspace{0.2cm} $2716.30/2596$ \hspace{0.2cm} & \hspace{0.2cm} $2716.10/2596$ \hspace{0.2cm} & \hspace{0.2cm} $2716.30/2596$ \hspace{0.2cm}\\
        & $=1.046$ & $=1.047$ & $=1.046$ & $=1.046$ & $=1.046$ & $=1.046$\\
        \hline\hline
    \end{tabular} }
\caption{Summary of best-fit values for the Kerr and non-Kerr models with \texttt{relxillion\_nk}. The ionization parameter $\xi_0$ is in units erg~cm~s$^{-1}$. The reported uncertainties correspond to the 90\% confidence level for one parameter ($\Delta\chi^2 = 2.71$). * indicates that the parameter is frozen in the fit. (P) indicates that the 90\% confidence level reaches the boundary of the parameter range.  \label{tab:fit-relxillion_nk}}
\end{table*}

\section{Discussion and conclusions \label{s-dis}}

In Paper~I, we constrained the KRZ deformation parameters by analyzing simultaneous \textsl{NuSTAR} and \textsl{XMM-Newton} observations of the supermassive BH in the galaxy MCG--06--30--15. In the present work, we have extended that study to constrain the KRZ deformation parameters with a stellar-mass BH. The constraints from stellar-mass and supermassive BHs can indeed test different physics because they probe different curvature regimes~\cite{Abdikamalov:2021zwv}. In some theories, new physics may be more easily discovered from the study of lighter BHs and in other models we may have the opposite case with larger deviations from the Kerr solution in heavier BHs.

All our results are perfectly consistent with the Kerr hypothesis. If we see the $\chi^2$ in Tab.~\ref{tab:fit-relxillion_nk}, the non-Kerr models can only marginally improve the fit with the Kerr background. The largest difference of $\chi^2$ is between the model with free $\delta_2$ and Kerr, where $\Delta\chi^2 = 1.16$. However, some parameters are poorly constrained. We recover the same trend found in Ref.~\cite{Nampalliwar:2019iti} and Paper~I: we can constrain $\delta_1$, $\delta_2$, and $\delta_5$, while we cannot constrain $\delta_3$ and $\delta_4$. As we see in Fig.~\ref{fig:contours}, the uncertainty on $\delta_3$ and $\delta_4$ is so large that even at a low confidence level we reach the upper and the lower boundaries of the parameter range. In Paper~I, we showed that even assuming simultaneous future observations with \textsl{Athena} and \textsl{eXTP} it is not possible to constrain these two deformation parameters. Their impact on the reflection spectrum is too weak and we need to consider other tests to constrain their value.

As discussed in the previous section, the quality of the fit improves if we use {\tt relxillion\_nk} and we leave the ionization index $\alpha_\xi$ free in the fit. In Ref.~\cite{Abdikamalov:2021rty}, we showed that the model with a non-trivial ionization gradient mainly improves the fit, but it does not have a significant impact on the estimate of most model parameters, in particular we did not see any difference in the estimate of the BH spin and the deformation parameters. However, this is not the case for the KRZ metric: if we assume a constant ionization parameter over all radii, we do not recover Kerr at 3-$\sigma$ in the models with free $\delta_4$ and $\delta_5$.

A natural question is how we can improve the constraints on the KRZ deformation parameters presented in this paper. Fig.~6 in Ref.~\cite{1869570} shows all the available constraints on the Johannsen deformation parameter $\alpha_{13}$ from stellar-mass BH data. Those constraints are obtained from different techniques and using X-ray and gravitational wave data. The constraints inferred from the 2019 \textsl{NuSTAR} observation of EXO~1846--031 are among the most stringent ones. Slightly stronger constraints can be obtained from three sources in which it is possible to combine the analysis of the reflection features with that of the thermal spectrum: GRS~1716--249~\cite{Zhang:2021ymo}, GRS~1915+105~\cite{1869570}, and GX~339--4~\cite{Tripathi:2020dni}. As discussed in Ref.~\cite{Tripathi:2020dni}, in those cases the systematic uncertainties become comparable with the statistical uncertainties, so better measurements can be obtained after improving our theoretical models; having higher quality data is not enough. 
For significantly better constraints from X-ray data, we presumably need to wait for the next generation of X-ray missions (e.g., \textsl{Athena} and \textsl{eXTP}) assuming the necessary development of the theoretical models and a better understanding of instrumental effects (e.g, calibration).

In Paper~III, we will present the constraints on the KRZ deformation parameters that can be inferred from the currently available \textsl{LIGO} and \textsl{Virgo} gravitational wave data of coalescing stellar-mass BHs. Some caution is necessary when we compare metric constraints from XRS and gravitational wave data within an agnostic framework such as the KRZ one, because the gravitational wave constraints are obtained assuming that the radiation-reaction force is prescribed as in general relativity~\cite{Cardenas-Avendano:2019zxd}. Under such an assumption, it turns out that XRS can constrain better some deformation parameters and gravitational wave data can constrain better other deformation parameters. This is perfectly understandable, because we go to analyze different relativistic effects, which are sensitive to different parts of the spacetime metric. We also note that XRS can essentially probe the spacetime up to the inner edge of the accretion disk, which is located at the ISCO and can be very close to the BH event horizon only for very fast-rotating objects, while we have no information from the region inside the ISCO. In the case of gravitational wave data, the signal of the ringdown phase can carry information about the presence of the event horizon~\cite{Maggio:2020jml}.


\vspace{0.5cm}

{\bf Acknowledgments --}
This work was supported by the Innovation Program of the Shanghai Municipal Education Commission, Grant No.~2019-01-07-00-07-E00035, the National Natural Science Foundation of China (NSFC), Grant No.~11973019, Fudan University, Grant No.~JIH1512604, and Fudan's Undergraduate Research Opportunities Program (FDUROP).
D.A. is supported through the Teach@T{\"u}bingen Fellowship. S.N. acknowledges support from the Alexander von Humboldt Foundation. The authors acknowledge support by the High Performance and Cloud Computing Group at the Zentrum f\"{u}r Datenverarbeitung of the University of T\"{u}bingen, the state of Baden-W\"{u}rttemberg through bwHPC and the German Research Foundation (DFG) through Grant No.~INST 37/935-1 FUGG.

\bibliography{references.bib}

\end{document}